\documentclass[12pt]{JHEP3}
\usepackage{epsfig}
\usepackage{amsmath}
\usepackage{amssymb}

\newcommand{\p}{\partial}
\title{Non-Extreme Black Holes from D-branes at Angles}

\author{Satoshi Nagaoka\\
High Energy Accelerator Research Organization (KEK)\\
Tsukuba, Ibaraki 305-0801, Japan\\
E-mail: \email{nagaoka@post.kek.jp}}

\abstract{
We construct the non-extreme solutions of non-orthogonal
intersecting D-branes.
The solutions reduce to non-extreme black holes upon the 
toroidal compactification.
We clarify the relation between two configurations 
with equal mass and charge,
one of which is non-orthogonal 
intersecting D-branes and the other one is orthogonal D-branes,
from supergravity and string theory perspective.
We also calculate mass and entropies for these black holes.
}


\preprint{KEK-TH-983, hep-th/0409137}

\begin{document}

\section{Introduction}

A microscopic understanding of the Bekenstein-Hawking 
entropy of extreme black holes is given 
in string theory \cite{SV}.
One of the directions of the extension of this result is 
to generalize away from extremality.
Non-extreme black holes play an important role in the study 
of the properties of realistic black holes. 
The non-extremality parameter $\mu$, which is the mass of neutral 
Schwarzschild black hole by setting all charges to zero,
interpolates between extremal and Schwarzschild black holes.
Entropies of non-extreme black holes are discussed in \cite{NEBH} 
in various dimensions.
$D=10$ type IIA supergravity can be obtained from the dimensional
reduction on a circle of $D=11$ dimensional supergravity.
The type IIA theory has the solitonic objects, D$p$-branes, with 
$p=0,2,4,6$, which preserve $1/2$ of a supersymmetry.
Orthogonal intersecting D-branes preserve part of a supersymmetry
and extreme black holes are constructed from these configurations.
Non-extreme generalization of orthogonal intersecting D-brane 
solutions of supergravity, which reduce to non-extreme black holes 
upon toroidal compactification, is shown in \cite{KT,CT}.
Non-orthogonal intersecting D-branes, which are widely studied 
as realistic brane models like Standard Model on intersecting 
D-branes recently,
 also preserve part of 
a supersymmetry \cite{BDL}\footnote{
Configurations of supersymmetric intersecting D-branes are constructed 
in many other papers, for example \cite{Jab}.
}.
Extremal black holes are constructed 
from branes at angles \cite{Angle,Angle2,CP}.
But it is not yet known the non-extreme generalization of non-orthogonal 
intersecting D-branes
\footnote{Recently, brane-antibrane systems at finite temperature
are analyzed to account for the entropy of the black branes 
far from extremality \cite{ddbar}.}.

In this paper, we will present the non-extreme solutions of 
supergravity from
2-angled non-orthogonal intersecting D-branes.
The relation between two configurations, one of which is non-orthogonal 
intersecting D-branes (A) and the other one is orthogonal D-branes (B),
is discussed from supergravity and string theory perspective.
In section 2, we construct the non-extreme solutions from 
intersecting D2-branes.
The correspondence between (A) and (B),
which is essential for constructing the non-extreme solution,
is discussed from supergravity.
In section 3, we compare (A) and (B) from string theory.
Section 4 is devoted to the conclusion and discussion.
In appendix, we calculate mass and entropies of non-extreme 
black holes which are obtained upon the toroidal compactification.

\section{Non-extreme solutions from branes at angles}

Part of type IIA supergravity Lagrangian 
which we need for the analysis
is written as
\begin{align}
L=\sqrt{-g} \left( e^{-2 \Phi }(R+4 (\nabla \Phi)^2) -\frac{1}{48} F_4^2
\right) \ ,
\end{align}
where the four-form field strength $F_4$ couples to D2-branes.
We adopt the string frame here.
Equations of motion for this action are written as
\begin{align} \notag
R&=-4 \nabla^2 \Phi +4 (\nabla \Phi)^2 \ , \\ \notag
R_{ij}& =-2 \nabla_i \nabla_j \Phi +\frac{1}{12} e^{2\Phi}
(F_i^{\ klm}F_{jklm}-\frac{1}{8} g_{ij} F_4^2) \ , \\
0&=\frac{1}{\sqrt{-g}} \p_i (\sqrt{-g} F^{ijkl}) \ .
\label{eom}
\end{align}
We consider two stacks of D2-branes 
and both stacks are
intersecting each other at angles $2 \theta$ 
to preserve 1/4 of a supersymmetry.
The location of the D2-branes is shown in Fig. 1.
For simplicity, we consider these configurations here,
but we can generalize the analysis to arbitrary numbers of 
stacks of intersecting D-branes.

\begin{figure}[htb]
\begin{center}
\begin{minipage}{12cm}
\begin{center}
\epsfxsize=12cm \leavevmode\epsfbox{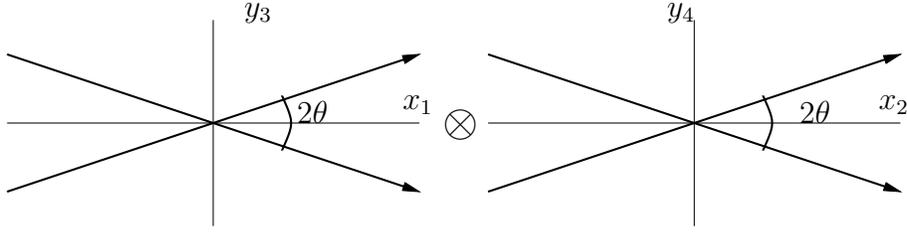} 
\put(-190,45){$x_1$}
\put(-10,45){$x_2$} 
\put(-90,80){$y_4$} 
\put(-250,80){$y_3$}
\put(-230,40){$2\theta$}
\put(-40,40){$2\theta$}
\put(-175,35){\Large $\otimes$}
\caption{Intersecting D2-branes: the first brane is denoted by
upper arrows on the 
$x_1y_3$ plane and $x_2y_4$ plane, and the second brane by
lower arrows on the $x_1y_3$ and $x_2y_4$ planes. 
These branes do not extend to other transverse directions.
Both of them 
have the same D2-brane charge $Q$
with different orientations.}
\end{center}
\end{minipage}
\end{center}
\end{figure}

Now, we would like to construct the non-extreme solution of this system.
The idea is to achieve understanding of the properties of the 
Schwarzschild black hole by doing perturbation theory in $\mu$.
An algorithm which leads to a non-extreme solution from a given
extreme solution is developed in \cite{CT}.
This procedure is applied for orthogonal intersecting D-branes
 \cite{CT}, but not yet for non-orthogonal intersecting D-branes.

\vspace*{0.5cm}
\noindent
\underline{Relation to the orthogonal intersecting D-branes} 
\vspace*{0.2cm}

Let us focus on $q$ D2-branes intersecting at two angles (Fig. 2 (A)).
The relation between $Q$ and $q$ is 
\begin{align}
Q\equiv q \cdot (\text{D2 charge per unit area}) \ .
\end{align}
We compactify $x_1,x_2, y_3$ and $y_4$ directions 
with the periods $a_1,a_2,b_1$ and $b_2$.
This system has topological charge 
\begin{align}
q(1,1) \otimes (1,1) \oplus q(1,-1) \otimes (1,-1)
=2q(1,0) \otimes (1,0) \oplus 2q(0,1) \otimes (0,1) \ ,
\end{align}
where $(q_1,q_3) \otimes (q_2,q_4)$ denotes that branes wrap 
$q_i$ times along $x_i(y_i)$ directions.

The system (B) also has topological charge 
$2q(1,0) \otimes (1,0) \oplus 2q(0,1) \otimes (0,1)$.

\begin{figure}[htb]
\begin{center}
\begin{minipage}{12cm}
\begin{center}
\epsfxsize=12cm \leavevmode
\epsfbox{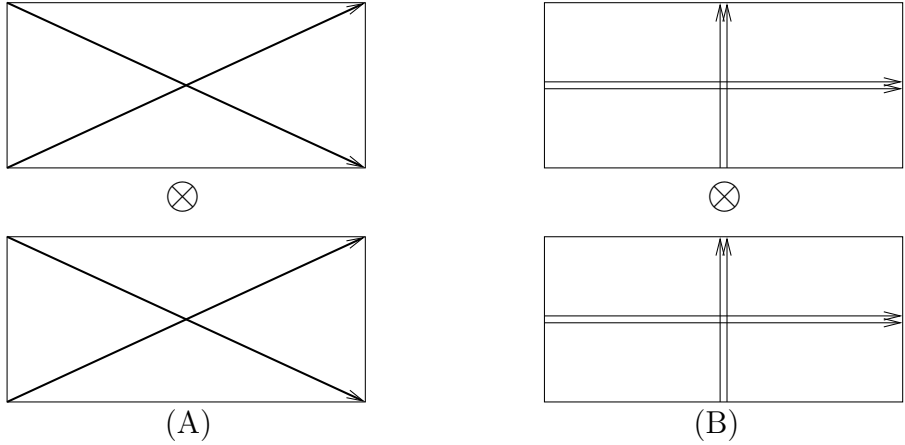} 
\put(-280,75){\Large $\otimes$}
\put(-75,75){\Large $\otimes$}
\put(-280,-10){(A)}
\put(-80,-10){(B)}
\caption{Two cycles are wrapped on $T^4$. Configuration (A) is 
non-orthogonal intersecting D-branes, on the other hand,  
configuration (B) is orthogonal intersecting D-branes. 
Both systems have 1/4 of 
a supersymmetry and topological charge $2q(1,0) \otimes (1,0)
 \oplus 2q(0,1) \otimes (0,1)$. }
\end{center}
\end{minipage}
\end{center}
\end{figure}

Next, we notice energy(tension) of these configurations.
Total energy of configuration (A) is
\begin{align}
E_{\text{A}}=2 c \sqrt{a_1^2+b_1^2} \sqrt{a_2^2+b_2^2} \ ,
\end{align}
where $c\equiv \frac{1}{g (2\pi)^2 (\alpha')^{3/2} }$ is D2-brane
tension and $g$ is string coupling constant.
On the other hand, total energy of configuration (B) is
\begin{align}
E_{\text{B}}=2 c (a_1a_2+b_1b_2) \ .
\end{align}
Using the relation which corresponds to the supersymmetric condition,
\begin{align}
\frac{b_1}{a_1}=\frac{b_2}{a_2} \ ,
\end{align}
we can easily check $E_{\text{A}}=E_{\text{B}}$.

Therefore, we conclude that there are supersymmetric 
2-angled non-orthogonal branes (A) which have the same 
total charge and tension as the orthogonal intersecting 
branes (B).
An non-extremalization procedure for the orthogonal
intersecting branes is already known \cite{CT}, then,
we can construct the non-extreme non-orthogonal intersecting D-brane
solutions by applying this procedure to the configuration (B).
It is interesting to construct supersymmetric 3-angled D-brane 
solutions which preserve $1/8$ of a supersymmetry.

Now, what is the difference between the configurations (A)
and (B)?
We can not find the difference between (A) and (B)
under the supergravity description because 
supergravity solution is constructed by the global charge,
which (A) and (B) have equally. On the other hand,
the mass spacing of the open strings connecting 
the different D-branes is determined by the 
intersection angle, therefore, configurations (A) and (B) 
have different mass spacing of the spectrum
\footnote{I would like to thank H. Shimada for bringing us 
this point.}.
The relations between (A) and (B) will be discussed further 
in section 3.

\vspace*{0.5cm}
\noindent
\underline{Constructing the solutions}
\vspace*{0.2cm}

We will construct the non-extreme solutions from 
non-orthogonal intersecting D-branes.
Extremal supergravity solutions of 
orthogonal intersecting D-branes are written as 
\begin{align}\notag
ds^2&=F^{1/2}\Big[
F^{-1} \left( - dt^2 +(1+X_1) (dx_1^2+dx_2^2) +
(1+X_2) (dy_3^2+dy_4^2) 
\right) \\ \notag
&+ \sum_{i=5}^9 dx_i^2
\Big] \ , \\ \notag
A_3&=dt \wedge (-\frac{X_1}{1+X_1}dx^1 \wedge dx^2+\frac{X_2}{1+X_2}
 dy^3 \wedge dy^4) \ , \\
e^{2 \Phi}&= F^{1/2} \ , \label{ext}
\end{align}
where
\begin{align} \notag
F&=\Pi_{i=1,2} F_i, \quad F_i=1+X_i \ , \\
X_1&=\frac{Q_1}{r^3}, \quad X_2=\frac{Q_2}{r^3} \ .
\end{align}
$Q_1$ and $Q_2$ are D2-brane charges along $x_1x_2$ plane and
$y_3y_4$ plane respectively.
Both configurations (A) and (B) have total charge $2q(1,0) \otimes (1,0)
 \oplus 2q(0,1) \otimes (0,1)$, which corresponds to 
\begin{align}  \notag
Q_1 &= \frac{2Qb_1b_2}{\sqrt{(a_1^2+b_1^2)(a_2^2+b_2^2)}}
=2Q \sin ^2 \theta \ , \\
Q_2 &= \frac{2Qa_1a_2}{\sqrt{(a_1^2+b_1^2)(a_2^2+b_2^2)}}
=2Q \cos ^2 \theta \ . \label{cha}
\end{align}
The solution (\ref{ext}) with charge 
\footnote{
$Q_1$ and $Q_2$ might become irrational number for some value $\theta$,
but this is no problem because only the ratio of the area
becomes irrational and the number of branes 
remains to be integer.
\begin{align}\notag
Q_1&=2q \sin^2 \theta \cdot (\text{D2 charge per unit area})  \ , \\
Q_2&=2q \cos^2 \theta \cdot (\text{D2 charge per unit area})  \ .
\end{align}
}
(\ref{cha}) 
is already found in \cite{Angle}.
Non-extreme procedure for orthogonal intersecting D-branes consists 
of the steps as
\begin{align} \notag
dt^2 &\to f(r) dt^2  \ , \\ \notag
\sum_{i=5}^9 dx_i^2 &\to f^{-1} dr^2 +r^2 d \Omega^2_4 \ , \\ \notag
F_i&\to \tilde{F}_i=1+\tilde{X}_i=1+\frac{\tilde{Q}_i}{r^3} \ , \\
A_3&\to \tilde{A}_3
=dt \wedge (-\frac{X_1}{1+\tilde{X}_1}dx^1 \wedge dx^2
+\frac{X_2}{1+\tilde{X}_2}
 dy^3 \wedge dy^4) , 
\end{align}
where
\begin{align}
f(r)=1-\frac{\mu}{r^3}  \ , 
\end{align}
and
\begin{align}
\tilde{Q}_i=\mu \sinh^2 \delta_i, \quad Q_i =\mu \sinh \delta_i
\cosh \delta_i \ , \quad i=1,2 \ .
\end{align}
The extremal limit corresponds to $\mu \to 0$ and $\delta_i \to 0$
with $Q_i$ kept fixed.
A non-extreme solution of the non-orthogonal intersecting D-branes 
is obtained as
\begin{align} \notag
ds^2&={\tilde{F}}^{1/2}\Big[
{\tilde{F}}^{-1} \left( - f dt^2 +(1+\tilde{X}_1) (dx_1^2+dx_2^2) +
(1+\tilde{X}_2) (dy_3^2+dy_4^2) 
\right)  \\ \notag
&+f^{-1} dr^2 +r^2 d\Omega_4
\Big] \ , \\ \notag
A_3&=\tilde{F}^{-1}dt \wedge (-X_1(1+\tilde{X}_2)dx^1
 \wedge dx^2+X_2(1+\tilde{X}_1) dy^3 \wedge
 dy^4) \ , \\
e^{2 \Phi}&=\tilde{F}^{1/2} \ , 
\end{align}
where
\begin{align}
\tilde{F}=\tilde{F}_1\tilde{F}_2=(1+ \tilde{X}_1)(1+\tilde{X}_2) \ ,
\end{align}
and the functions $\tilde{X}_1$ and $\tilde{X}_2$ are harmonic functions 
in the transverse space,
\begin{align} \notag
\tilde{X}_1= \frac{\tilde{Q}_1}{r^{3}}, \quad
\tilde{Q}_1=-\frac{\mu}{2}+\sqrt{(2 Q \sin^2 \theta)^2+
(\frac{\mu}{2})^2} \ , \\
\tilde{X}_2=\frac{\tilde{Q}_2}{r^{3}} , \quad
\tilde{Q}_2=-\frac{\mu}{2}+\sqrt{(2 Q \cos^2 \theta)^2+
(\frac{\mu}{2})^2} \ .
\end{align}
This solution satisfies the equations 
of motion (\ref{eom}).
In the extremal limit $\mu \to 0$, we obtain the extreme solution 
(\ref{ext}).

\section{State counting of D-branes}

We consider the number of states of configurations (A) and (B) here.
The massless degrees of freedom which contribute to the entropy
formula are associated with open strings stretching the intersecting
D-branes \cite{Pol}.
Configuration (B) is constructed from 
$2q$ D2-branes
along $x_1x_2$ plane and $2q$ D2-branes along $y_3y_4$ plane.
Therefore, the configuration (B) has $4q^2$ massless excitation modes
which contribute to the entropy.

The number of massless modes of non-orthogonal intersecting D-branes (A)
which contribute to the entropy formula 
is obtained in \cite{CP}.
We will briefly review the bosonic part of the analysis.
Fundamental strings stretching two intersecting D-branes are
described by the following boundary condition,
\begin{align} \notag
&\p_\sigma Z_a +i \tan \theta \p_\tau Z_i |_{\sigma=0}=0, \\
&\p_\sigma Z_a -i \tan \theta \p_\tau Z_i |_{\sigma=\pi}=0 \ ,
\end{align}
where $Z_a=x_a+i y_{a+2}$ for $a=1,2$. The classical solutions 
of the equations of motion for the complex bosons with these 
boundary conditions have mode expansions which are written as 
\begin{align}
Z_a=z_a+i(\sum_{n=1}^\infty a_{n-\epsilon}^a \phi_{n-\epsilon}
(\tau,\sigma)-\sum_{n=0}^\infty a_{n-\epsilon}^{a\dagger}
 \phi_{-n-\epsilon}(\tau,\sigma) ) \ ,
\end{align}
where 
\begin{align}
\phi_{n-\epsilon}=\frac{1}{\sqrt{|n-\epsilon|}}\cos ((n-\epsilon)\sigma
+\theta) \exp (-i (n-\epsilon )\tau) \ , 
\end{align}
with $\epsilon\equiv \frac{2\theta}{\pi}$.
The commutators of zero modes $x_a$ and $y_{a+2}$ are 
obtained as
\begin{align}
[x_1,y_3]=[x_2,y_4]=i\frac{\pi}{2\tan \theta} \ .
\end{align}
Dirac quantization condition restricts the values of $x_1$ and $x_2$
as
\begin{align} \notag
x_1&=\frac{n_1p_1p_3a_1}{N_1}  \ , \quad n_1: \text{integer} \ ,
\\ \notag
x_2&=\frac{n_2p_2p_4a_2}{N_2} \ , \quad n_2: 
\text{integer} \ , \\
N_1&= |p_1q_3-p_3q_1| \ , \quad
N_2=|p_2q_4-p_4q_2| \ ,
\end{align}
where $p_i$ and $q_i$ ($i=1,2,3,4$) are wrapping numbers on cycles
$x_i(y_i)$ of the first and second D2-branes.
$N_1$ and $N_2$ denote the intersection numbers of $x_1y_3$ plane and
$x_2y_4$ plane respectively.
The degeneracy of the massless mode is 
$N_1N_2$.
The wrapping number of configuration (A) is
\begin{align} \notag
(p_1,p_3) \otimes (p_2,p_4)&=(1,1)\otimes(1,1) \ , \\
(q_1,q_3) \otimes (q_2,q_4)&=(1,-1)\otimes(1,-1) \ ,
\end{align}
on each $T^2 \times T^2$,
therefore, the total number of massless mode is obtained as
\begin{align}
q^2N_1N_2=q^2|(-1-1)(-1-1)|=4q^2 \ ,
\end{align}
which is equal to the number of massless modes of (B).

On the other hand,
the mass spectrum of open string ending on D-branes is
written as \cite{BDL}
\begin{align}
m^2=\frac{2n\theta}{\pi \alpha'} \ , \quad n=0,1,\cdots \ ,
\end{align}
therefore, configuration (A) and (B) have different mass spectrum.

\section{Conclusion and discussion}

It is well-known that
the orthogonal intersecting branes can be generalized to the 
non-extreme solutions, on the other hand,
it was not yet known the non-extremalization 
procedure of the non-orthogonal
intersecting D-branes. In this paper, we have
constructed the non-extreme black holes from
non-orthogonal intersecting D-branes.
The essential point for the construction
is that we can transform the basis of the brane charge
from non-orthogonal D-branes to orthogonal D-branes.
After the transformation, we non-extremalize the solutions for each 
orthogonal intersecting brane charges.

The configuration (A) and (B) have the same mass and 
global charge, therefore,  
(A) and (B) reduce to the same black hole.
They also have the same number of massless states of D-branes.
But
they are different configurations in string theory, that is, 
they have different mass spectrum. In other words,
this is something like `hair' of the black hole.

Intersecting D-branes are the fundamental and important objects 
in string theory, and it has much possibility to bring  
information from the string theory to the realistic world.
In our research directions,
first, it is interesting to construct the 3-angled intersecting D-brane 
solutions, which is the geometrical aspects 
to interpolate between the string theory and the realistic world.
In these configurations, we can consider $D=4$ black holes, 
and black holes with more than 3-charges.
Another context is to realize 
Standard Model on intersecting D6-branes, which are constructed as
 3-cycles wrapped 
on $(T^2)^3$ with 3 intersection angles.
When we consider M-theory lift of intersecting D6-branes 
(plus O6-planes), the background is a singular 7-manifold with $G_2$
holonomy.
It might be interesting to analyze it from this direction.
Second, to clarify the relation between supergravity approach and 
the effective field theory (gauge theory) approach is also interesting.
In \cite{YM}, intersecting branes were analyzed by using
Yang-Mills theory in the small intersection angles.
We have extended the angle parameters to arbitrary value for the
non-extreme black brane solutions.
It might become some clue to explain the recombination mechanism,
which corresponds to Higgs mechanism on Standard Model, 
on the intersecting D-branes from the supergravity approach. 
Third, the black hole degrees of freedom should be studied 
further. It might be interesting to consider the 
`hair' conjectured by Mathur \cite{MSS},
which is closely related to the information paradox, in this case.

\section*{Acknowledgements}

I would like to thank H. Fuji, K. Hotta, Y. Hyakutake, Y. Sekino,
M. Shigemori, H. Shimada, T. Takayanagi and 
especially Y. Kitazawa 
for quite useful discussions and comments.
I would also like to thank Y. Kitazawa for carefully 
reading this manuscript.

\appendix

\section{Mass and entropies of non-extreme solution}

The area of the horizon at $ r=\mu^{1/3}$ is
\begin{align}\notag
A_8&=[\frac{8 \pi^2}{3} a_1a_2b_1b_2 r^4 \tilde{F}^{1/2} ]_{r=\mu^{1/3}}
 \\
&=\frac{8 \pi^2}{3} a_1a_2b_1b_2 \mu^{4/3} \cosh \delta_1 \cosh \delta_2 
 \ ,
\end{align}
where
internal coordinates $x^a (y^a)$ have period $a_1,a_2,b_1$ and $b_2$.
The corresponding 6-dimensional metric is 
\begin{align}
ds_6^2=-\tilde{F}^{-1/2} f dt^2+\tilde{F}^{1/2}
[f^{-1}dr^2+r^2 d\Omega_4^2] \ .
\end{align}
The ADM mass is calculated as
\begin{align} \notag
M&=\frac{4\pi^2 a_1a_2b_1b_2}{3 \kappa^2} 
(4 \mu +3 \tilde{Q}_1 +3 \tilde{Q}_2) \\
 &=\frac{4 \pi^2a_1a_2b_1b_2}{\kappa^2}
\left(\sqrt{(2 Q \cos^2 \theta)^2+(\frac{\mu}{2})^2} 
+\sqrt{(2 Q \sin^2 \theta)^2+(\frac{\mu}{2})^2}+\frac{\mu}{3}\right) \ ,
\end{align}
where $\kappa $ is 6-dimensional Planck constant.
In the extremal limit, we obtain
\begin{align}
M=\frac{8 \pi^2a_1a_2b_1b_2 Q}{\kappa^2} \ .
\end{align}
Bekenstein-Hawking entropy is 
\begin{align} \notag
S_{\text{BH}}&=\frac{2 \pi A_8}{\kappa^2} \\ \notag
&=\frac{16\pi^3 a_1a_2b_1b_2}{3 \kappa^2}\mu^{4/3} \cosh \delta_1
\cosh \delta_2 \\
&=\frac{16 \pi^3 a_1a_2b_1b_2}{3 \kappa^2} \mu^{1/3}\left(
\sqrt{(2 Q \sin^2 \theta)^2+(\frac{\mu}{2})^2}+\frac{\mu}{2}
\right)^{\frac{1}{2}}
\left(
\sqrt{(2 Q \cos^2 \theta)^2+(\frac{\mu}{2})^2}+\frac{\mu}{2}
\right)^{\frac{1}{2}} \ ,
\end{align}
and Hawking temperature is 
\begin{align}
T=\frac{3 \mu^{2/3}}{4 \pi} 
\left(
\sqrt{(2 Q \sin^2 \theta)^2+(\frac{\mu}{2})^2}+\frac{\mu}{2}
\right)^{-\frac{1}{2}}
\left(
\sqrt{(2 Q \cos^2 \theta)^2+(\frac{\mu}{2})^2}+\frac{\mu}{2}
\right)^{-\frac{1}{2}} \ .
\end{align}
In the near extremal limit, we obtain 
\begin{align}\notag
M&=M_0 +\Delta M +{\cal O} (\mu^2) \ , \\
M_0&=\frac{4\pi^2 a_1a_2b_1b_2 Q}{ \kappa^2}, \quad
\Delta M=\frac{4\pi^2 a_1a_2b_1b_2 \mu}{3 \kappa^2} \ ,
\end{align}
and 
\begin{align} \notag
S_{\text{BH}}&=\frac{16 \pi^3 a_1a_2b_1b_2 }{3 \kappa^2} \mu^{1/3}
Q\sin 2 \theta\left(1+\frac{\mu}{2 Q \sin^2 2 \theta }+
{\cal O} (\mu^2)\right) \\
&\sim (\frac{4\pi^2a_1a_2b_1b_2}{3\kappa^2})^{2/3} 4\pi Q \sin 2\theta \ .
\end{align}
By using the Hawking temperature, $S_{\text{BH}}$ is written as
\begin{align}
S_{\text{BH}}=\frac{4 \pi^2 a_1a_2b_1b_2}{\kappa^2}(\frac{4 \pi}{3})^{3/2}
(Q\sin2 \theta)^{3/2} T^{1/2} \ ,
\end{align}
where
\begin{align}
T=\frac{3 \mu^{2/3}}{4 \pi} \left(
\frac{1}{Q \sin 2 \theta}
\right) \ .
\end{align}

%%%%%%%%%% References %%%%%%%%%%%%%%%%%%%%%%%%%
\newcommand{\J}[4]{{\sl #1} {\bf #2} (#3) #4}
\newcommand{\andJ}[3]{{\bf #1} (#2) #3}
\newcommand{\AP}{Ann.\ Phys.\ (N.Y.)}
\newcommand{\MPL}{Mod.\ Phys.\ Lett.}
\newcommand{\NP}{Nucl.\ Phys.}
\newcommand{\PL}{Phys.\ Lett.}
\newcommand{\PR}{ Phys.\ Rev.}
\newcommand{\PRL}{Phys.\ Rev.\ Lett.}
\newcommand{\PTP}{Prog.\ Theor.\ Phys.}
\newcommand{\hep}[1]{{\tt hep-th/{#1}}}
%%%%%%%%%%%%%%%%%%%%%%%%%%%%%%%%%%%%%%%%%%%%%%%

\end{document}